# Single molecule enzymology à la Michaelis-Menten

Ramon Grima[1], Nils G. Walter[2] and Santiago Schnell[3*]

[1] School of Biological Sciences and SynthSys, University of Edinburgh, Edinburgh, UK

[2] Department of Chemistry, University of Michigan, Ann Arbor, Michigan, USA

[3] Department of Molecular & Integrative Physiology, Department of Computational Medicine & Bioinformatics, and Brehm Center for Diabetes Research, University of Michigan Medical School, Ann Arbor, Michigan, USA

\* To whom the correspondence should be addressed. E-mail: schnells@umich.edu

## Abstract

In the past one hundred years, deterministic rate equations have been successfully used to infer enzyme-catalysed reaction mechanisms and to estimate rate constants from reaction kinetics experiments conducted *in vitro*. In recent years, sophisticated experimental techniques have been developed that allow the measurement of enzyme-catalysed and other biopolymer-mediated reactions inside single cells at the single molecule level. Time course data obtained by these methods are considerably noisy because molecule numbers within cells are typically quite small. As a consequence, the interpretation and analysis of single cell data requires stochastic methods, rather than deterministic rate equations. Here we concisely review both experimental and theoretical techniques which enable single molecule analysis with particular emphasis on the major developments in the field of theoretical stochastic enzyme kinetics, from its inception in the mid-twentieth century to its modern day status. We discuss the differences between stochastic and deterministic rate equation models, how these depend on enzyme molecule numbers and substrate inflow into the reaction compartment and how estimation of rate constants from single cell data is possible using recently developed stochastic approaches.

**Abbreviations**

DRE, deterministic rate equation; FCS, fluorescence correlation spectroscopy; PSF, point spread function; FRET, fluorescence resonance energy transfer; CME, chemical master equation; SSA, stochastic simulation algorithm; EMRE, effective mesoscopic rate equation.

## Introduction

For just over a century, enzymologists have endeavoured to infer the molecular mechanism and estimate kinetics constants of enzyme-catalysed reactions using four experimental approaches: initial rate, progress curve, transient kinetics and relaxation experiments [1, 2]. The mechanistic basis of the simplest single enzyme, single substrate reaction was proposed by Victor Henri in 1902 [3-5]. This reaction mechanism of enzyme action consists of a reversible step between an enzyme E and a substrate S, yielding the enzyme-substrate complex C, which subsequently forms the product P:

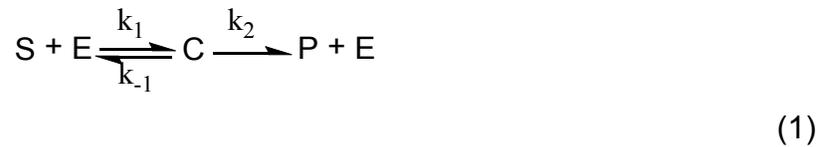

(1)

where $k_1$, $k_{-1}$ and $k_2$ are the rate constants of the reaction. Eq. (1) is known as the Michaelis and Menten reaction mechanism of enzyme action, because Leonor Michaelis and Maud Leonora Menten showed a century ago [6] that enzymes can be investigated by measuring the initial rate of product formation under certain experimental conditions [2, 7]. The initial rate of production formation $(v_0)$ is given by the Michaelis-Menten equation:

$$v_0 = \frac{dp}{dt} = \frac{k_2 e_0 s}{K_M + s}.$$

(2)

In the above expression, $k_2$ is the turnover number, $e_0$ is the initial enzyme concentration in the experiment, $K_M = (k_{-1} + k_2)/k_1$ is the Michaelis-Menten constant and $s$ is the initial substrate concentration. $v_0$ is a rectangular hyperbolic function of $s$, which increases rapidly until it reaches the saturating value of the limiting rate, $V = k_2 e_0$, at high $s$. The simple saturating function of the Michaelis-Menten equation has been a cornerstone of enzyme kinetics ever since, because it allows the estimation of the kinetic parameters characterising the enzymatic catalysis, $V$ and $K_M$, from measurements of the initial rate of product formation under different substrate concentrations in quasi-steady state conditions (see [7] for a review). With the advent of computers, nowadays kinetic parameters are generally estimated from time course experiments by numerically integrating the reaction rate expressions [8-10].

Michaelis's and Menten's lasting contribution to enzymology has played a fundamental role in understanding enzyme biochemistry in the test tube. The Michaelis-Menten equation is a deterministic rate equation (DRE), which implicitly assumes that the number of enzyme and substrate molecules is macroscopically large [11, 12]. This is a fundamentally limiting assumption when one considers that the number of molecules of many chemical species inside cells range from tens to a few thousands [13, 14], a number many orders of magnitude smaller than that in typical test tube experiments. In low molecule number conditions, time-course measurements are not smooth but are rather characterised by large fluctuations (see, **FIG. 1B**). This intrinsic noise stems from the random timing of biochemical reaction events. The randomness has various sources of origin including the Brownian motion of reactants [15]. The noise in the concentration of a given molecular species roughly scales as the inverse square root of the total number of molecules of the species [16]. This implies that stochastic fluctuations are always present, but they are irrelevant in bulk conditions. For this reason, DREs describe well reaction dynamics when molecules are present in large numbers. By the same reasoning, however, DREs like the Michaelis-Menten equation cannot be used to investigate noisy intracellular or single molecule enzyme catalysed reactions.

During the last twenty years, the development of mathematical and computational approaches to investigate the inherent stochasticity of reactions inside the cell has been propelled by advances in experimental techniques that are capable of following reactions at the single molecule level using fluorescence microscopy and related optical methods [14, 17-24]. Here, we first survey some of the improvements in single molecule analysis developed to investigate intracellular reactions. Second, we present the major developments in the field of theoretical stochastic enzyme kinetics – from its inception in the mid-twentieth century until today – that deal with the resulting data. Our aims are to:

(1) Highlight the differences and similarities between stochastic and deterministic rate equations. (2) Discuss the differences between stochastic models of enzyme kinetics in a closed compartment and in a compartment with substrate inflow. (3) Clarify how the kinetic parameters can be estimated from single molecule data and how the reliability of estimation depends on the choice of modelling framework. (4) Stress that only a small number of the theoretical predictions have been verified by experiments and hence single molecule enzymology still presents many exciting challenges.

## Single molecule analysis in real-time

In 1959 Richard Feynman first predicted that "there's plenty of room at the bottom" [25], and since then the quest to detect and manipulate fewer and fewer molecules in ever smaller volumes has begun to make rapid strides [14]. Two main types of microscopy approaches for directly observing the behaviour of single molecules have emerged – optical detection, largely through measurement of a fluorescence signal, but also through measurement of absorption or scattering; and mechanical detection such as the topological mapping by atomic force microscopy or the application of controlled molecular scale forces [14, 26]. For details, the reader is referred to some of the many recent reviews on the topic [14, 17-24]. Briefly, single molecule approaches can detect classic enzymatic substrate turnover, but also other biopolymer-mediated reactions such as binding and dissociation events and conformational changes. Often, the observation of multiple events from a single enzyme or biopolymer (such as RNA, DNA or a polysaccharide) lends increased statistical significance to the signal, improving the ability to distinguish from spurious background events (such as detection noise and non-specific binding) at the relatively low signal-to-noise ratios of single molecule detection. Fundamentally, if the molecule of interest can be immobilized to be observed for an extended period of time and/or act as prey to capture a diffusing "predator" molecule (or substrate) [27-29], the likelihood of observing multiple events and the signal-to-noise ratio both increase, and the confidence in interpretation of the data rises. For force-based techniques, immobilization on a solid support is essential for providing a topological map across the support matrix and for applying a known force. However, certain sensitive fluorescence detection techniques can also be applied to freely diffusing molecules, as long as averaging over multiple molecules (rather than multiple events from a single molecule) yields useful information such as the intracellular diffusion constant in fluorescence correlation spectroscopy (FCS) [22, 30]. Conversely, if only the number of molecules needs to be counted or their position recorded, diffusion or photobleaching to remove those already detected can be beneficial [14, 20, 24, 29].

One of the major advantages of single molecule fluorescence techniques in particular is that they can be applied readily to measuring the relatively unperturbed real-time behaviour of single molecules inside live cells. Since the number of identical biopolymer molecules in a single cell typically ranges from just 1 to ~1,000 [14, 31] or, in some cases like the ribosome, several 10,000s, and since the volume of the cell is small (eukaryotes typically have a diameter of 10-100 µm, and bacteria 0.2-2 µm), microscopic detection of those few single molecules becomes critical. Conversely, single molecule fluorescence microscopy benefits from low molecule numbers since each molecule will then appear as a signal (termed a point spread function, or PSF) that is spatially resolved from others. In fact, photo-activation and -switching are used as "tricks" to only turn on sparse numbers of molecules at each detection time point to prevent excessive molecule numbers and poorly resolved signals when imaging a larger field of view in a wide-field microscope [14]. Additionally, the limited focal depth of ~500 nm of a high numerical aperture microscope objective effectively removes molecules outside of the imaging plane through defocused blurring. Alternatively, confocal fluorescence microscopy (as in FCS) and spatial confinement techniques have been developed to detect isolated single molecules from small volume elements using a point detector [22, 30].

When applied to the cell interior, the main observables of these fluorescence microscopy techniques are [14]:

- The location of the fluorophore labelled molecule, which nowadays is determined routinely at ~10- to 20-fold higher resolution than foreseen by the classical Abbe law or Rayleigh resolution limit of ~λ/2 (where λ is the wavelength of light used for imaging, typically around 450-700 nm), based on either software-fitting or optically shaping the PSF;

- The brightness of the fluorophore labelled molecule, which under certain circumstances can reveal the stoichiometry of a multi-molecule complex, either through careful calibration or through counting of stochastic, stepwise photobleaching events, where each step corresponds to the loss of signal from one fluorophore; and

- The spectral properties (colour) of the fluorophore label; the spectral resolution when detecting single molecules is limited due to the limited number of photons emitted, but can for example resolve the relative contribution of the red-shifted so-called acceptor to the fluorescence signal of a donor-acceptor doubly labelled molecule. This feature has powerfully been used to measure the distance between the attachment sites of a donor and acceptor that undergo distance-dependent fluorescence resonance energy transfer (FRET) [32].

What makes intracellular fluorescence microscopy particularly powerful is that it offers spatiotemporal resolution, that is, changes in any of the observables can be monitored in real-time coupled with location information (see, **FIG. 1**). This resolution directly yields kinetic information, for example, when measuring: temporal changes in molecule location to assess diffusion coefficients [33]; assembly stoichiometry or fluorogenic substrate turnover through stepwise changes in brightness that may systematically vary with time as observed through time-lapse experiments [33]; or temporal changes in molecule conformation or configuration when observing changes in FRET between a judiciously placed donor-acceptor pair [32, 34]. These kinetic data may additionally reflect intracellular reactions such as binding and dissociation (for example, when two molecules begin and cease, respectively, to diffuse or localize together) or substrate turnover by an enzyme (for example, when such a turnover is associated with the appearance or disappearance of a fluorescence signal or a change in FRET). Intracellular single molecule fluorescence techniques in particular have therefore fuelled the need for analysing (and further developing) stochastic reaction kinetics as detailed next sections.

## Physicochemical theory of stochastic reaction kinetics

To understand how to model stochastic chemical reactions, let us consider a hypothetical setup consisting of a large number of independent samples of the same chemical reaction each with identical initial conditions. Due to the inherent stochasticity of chemical interactions, the number of molecules at a given fixed time varies from sample to sample. This variation is captured by the fraction (or probability) of samples at time $t$, $P(n_1, n_2,..., t)$, containing $n_1$ number of molecules of species 1, $n_2$ number of molecules of species 2, etc. The stochastic description of the reaction kinetics then is given by a differential equation for this probability; this is in contrast to DREs, which are differential equations for the mean concentrations. Over the years, this probabilistic approach was developed to model any set of elementary reaction steps. It is known nowadays as the chemical master equation (CME). The CME can be derived from simple laws of probability and microscopic physics [35]. Its microscopic validity has also been tested and verified by molecular dynamics [36, 37] for dilute chemical systems and using Brownian dynamics simulations [38] for non-dilute crowded systems. The major assumption underlying the CME is that reactions are occurring in well-mixed environments, which is also an assumption intrinsic to DRE models. Typically, the well-mixed reaction environment assumption is satisfied in sub-micron intracellular compartments since normal diffusion creates homogeneity of molecular species over small volumes.

DREs can be obtained from the CMEs in the macroscopic limit, i.e., the limit of large volumes at constant concentration (which implies the limit of large molecule numbers) [16, 39]. Thus the CME approach is more fundamental then the DRE approach. The two approaches will generally lead to different predictions for the mean concentrations [11, 40] and hence one should interpret results obtained using DREs with caution, relying on them only when the molecule numbers are quite large.

To estimate the mean concentrations of the CME model, we need to study the first moments of the probability distribution of the CME. Higher-order moments of the CME probability distribution present information about fluctuations which are not available in DREs. The second-order moment is an illustrative example: it describes the variance of fluctuations about the mean concentrations providing a measurement of the variability between independent experimental realisations of the chemical reaction under study. Recent work suggests that accurate estimates of rate constants can be obtained by making use of both first and second moment information [41]. Such higher-order information could also be used to distinguish between rival mechanistic models, such as the Michaelis-Menten and Nuisance-Complex reaction mechanisms of enzyme action [5, 42].

The primary reason which has limited the exploitation of the CME approach is the lack of exact solutions. Hence, much of the literature to date has focused on identifying cases where exact solutions of the CME are possible and more generally on obtaining approximate solutions to the moments of the CME using sophisticated mathematical approaches. In what follows we review some of the major advances made in these directions, in particular focusing on the differences between stochastic kinetics in a closed compartment and in a compartment with substrate inflow (see, **FIG. 2**).

**Stochastic analysis of the Michaelis-Menten reaction mechanism**

The CME for the single enzyme, single substrate Michaelis-Menten reaction mechanism (1) was first derived and studied by Anthony F. Bartholomay in 1962 [15]. He introduced a time-evolution equation for the probability $P(n_S, n_E, n_C, n_P, t)$, where $n_S, n_E, n_C$ and $n_P$ are the molecule numbers of substrate, enzyme, enzyme-substrate complex, and product, respectively, at time $t$ in a closed compartment. Bartholomay's equation is:

$$\frac{dP(n_S,n_E,n_C,n_P,t)}{dt} = \frac{k_1}{\Omega}\left[(n_S+1)(n_E+1)P(n_S+1,n_E+1,n_C-1,n_P,t) - n_S n_E P(n_S,n_E,n_C,n_P,t)\right]$$
$$+ k_{-1}\left[(n_C+1)P(n_S-1,n_E-1,n_C+1,n_P,t) - n_C P(n_S,n_E,n_C,n_P,t)\right]$$
$$+ k_2\left[(n_C+1)P(n_S,n_E-1,n_C+1,n_P-1,t) - n_C P(n_S,n_E,n_C,n_P,t)\right],$$

(3)

where the terms on the three lines describe three steps: the association of enzyme and substrate, the breakdown of complex into substrate and enzyme and the breakdown of complex into enzyme and product, respectively. The parameter $\Omega$ is the volume of the compartment in which the reaction occurs. Note that the contribution of the three steps to the overall dynamics is regulated by the constants $k_1/\Omega$, $k_{-1}$ and $k_2$ which are the inverse timescales associated with each of the aforementioned steps. We note that the rate constants $k_1$, $k_{-1}$ and $k_2$ are precisely the same constants which appear in the DRE formulation of kinetics. A detailed explanation of the construction of CMEs is beyond the scope of this review; the reader is referred to more specialized reviews and books on this topic [12, 16, 43].

Bartholomay demonstrated that the CME of the Michaelis-Menten reaction mechanism (1) reduces to the DREs in the macroscopic limit of large molecule numbers. In particular the DREs are obtained from the CME by assuming that the covariance of fluctuations in the numbers of enzyme and substrate molecules is zero; this condition is only true in the limit of large molecule numbers since the size of fluctuations decreases with increasing molecule numbers [12]. In 1964 Jachimowski et al. [44] derived a CME model for the Michaelis-Menten reaction mechanism with competitive inhibition and for an enzyme with two alternative substrates. They also showed that the CMEs are equivalent to their DRE counterparts in the macroscopic limit.

In general, single molecule biophysicists and chemists investigating enzyme-catalysed and other biopolymer-mediated reactions are interested in the case where the molecule numbers are not very large. The question then is whether the CME can be solved exactly analytically for reactions characterised by a small number of molecules. This is the topic of the next section.

**Analysis of the Michaelis-Menten reaction catalysed by few enzyme molecules**

Aranyi and Toth [45] were the first to systematically study the CME introduced by Bartholomay. They considered the special case where there is only one enzyme molecule with several substrate molecules in a closed compartment (see, Fig. **2A**) and showed that the CME can then be solved exactly. The exact solution consists of the probability distribution of the state of the system at any time point. This is remarkable when one considers that it is impossible to solve the DREs without imposing restrictions on the reaction conditions such as pseudo first-order kinetics [46], or applying an approximation [47] such as quasi-steady state assumption [48], rapid-equilibrium assumption [49] or reactant stationary assumption [50].

From the exact solution of the probability distribution, they derived exact expressions for the time course of the mean substrate and enzyme concentrations and compared them with those obtained by numerical integration of the DREs. Interestingly, Aranyi and Toth [45] found differences of 20-30% between the average substrate concentrations calculated using the DREs and the CME for the same set of rate constants and for the case of one enzyme reacting with one substrate molecule (see **FIG. 3A**). If the initial number of substrate molecules is increased to 5 whilst keeping the same rate constants then one notices that the difference between the DRE and CME results becomes negligibly small (see **FIG. 3B**). Generally it can be shown that the discrepancy between the two approaches stems from the fact that the mean concentrations, in chemical systems involving second-order reactions, are dependent on the size of the fluctuations in a CME description and independent in a DRE description [51]. The discrepancies become smaller for larger numbers of substrate molecules since fluctuations roughly scale as the inverse square of the molecule numbers [12]. This important contribution by Aranyi and Toth went largely unnoticed at the time, because experimental approaches did not have the resolution for measuring single enzyme-catalysed experiments to test the theoretical results.

With the advent of single molecule experiments, the differences between deterministic and stochastic enzyme kinetics have begun to be explored in the last decade. In this context the inverse mean time between successive product formation events is equivalent to the mean rate of product formation; one can then ask what the dependence of this quantity is on the mean substrate concentration in a single enzyme experiment. By assuming that the substrate is much more abundant than the enzyme, Kou et al. [52] and Qian [53] simplified the CME governing the Michaelis-Menten reaction since the association of enzyme and substrate is effectively pseudo first-order during the initial transient of the reaction. They found that the relationship between the initial mean rate of product formation and the initial substrate concentration is given by the Michaelis-Menten equation (2). Their relationship is frequently termed the single molecule Michaelis-Menten equation. The theoretical predictions were confirmed using single molecule experiments monitoring long time traces of enzymatic turnovers for individual β-galactosidase molecules by detecting one fluorescent product at a time [29].

The discovery of the single molecule Michaelis-Menten equation is an interesting result. From the perspective of the DRE approach this can be seen as obvious finding since the single enzyme-many substrate molecule setup is the ultimate realization of a particular condition (not the general condition) under which the deterministic Michaelis-Menten equation is valid, namely that the initial substrate concentration greatly exceeds that of the initial enzyme concentration [7, 47]. However, note that this line of thinking does presume the correctness of the DRE approach even for small molecule numbers, which is clearly not the case generally. Hence from the latter perspective the derivation of a single molecule Michaelis-Menten equation is surprising.

Interestingly, Kou et al. [52] showed that if the states of the enzyme interchange between a number of interconverting conformations (dynamic disorder) then one obtains a slightly more complicated equation than the single molecule Michaelis-Menten equation (see Eq. (22) in [52] for the case of two conformations). Use of this equation to understand single-molecule data is warranted whenever one suspects dynamic disorder to be at play, namely when the distribution of times between successive product formation steps is multi-exponential [29]. Nonetheless it should be kept in mind that the deviations from the Michaelis-Menten equation are small for several cases of dynamic disorder, e.g. when fluctuations between conformer forms of the enzyme and enzyme-substrate intermediates occur on a much longer timescale than the turnover time [52], and hence the presence of dynamic disorder does not necessarily preclude the use of the single-molecule Michaelis-Menten equation.

In summary, taking together the results of Aranyi and Toth [45], Kou et al. [52] and Qian [53], we have an emerging theoretical picture of the differences between the DREs and CME description for the Michaelis-Menten type reaction catalysed by a single enzyme molecule. The DRE and CME approaches give virtually indistinguishable results for the temporal evolution of the mean substrate concentrations and for the initial rates of product formation whenever the initial number of substrate molecules is larger than a few molecules. These predictions have been confirmed by recent single molecule experiments; however the predicted discrepancies between CME and DRE approaches for the interaction of a single molecule of substrate and of enzyme still await experimental confirmation.

Of course generally it is unlikely that there is one single enzyme molecule inside a subcellular compartment or an experimental set up. However the single enzyme case is useful because it allows us to estimate the maximum deviations one would expect in typical scenarios. To date, it has not been possible to obtain an exact analytical expression for the probability distribution solution of the CME of the Michaelis-Menten reaction catalysed by many enzyme molecules. For small numbers of enzyme molecules one can solve the CME numerically using the finite state projection algorithm [54]. However, the most common method of probing the CME is the stochastic simulation algorithm (SSA), which is a Monte Carlo technique generating sample paths of the stochastic process described by the CME [43]. Using the SSA it has been shown that the differences between the mean concentrations predicted by DREs and the CME for enzyme molecules greater than a few tens and characterized by the condition $k_2 \ll k_{-1}$ in the Michaelis-Menten reaction mechanism (1), are very small and hence can typically be ignored [55-57].

## Stochastic analysis of the Michaelis-Menten reaction mechanism with substrate inflow

Thus far, we have considered the Michaelis-Menten reaction mechanism (1) in a closed compartment (**FIG. 2A**). This mechanism ignores the fact that under physiological conditions substrate is synthesised by upstream processes and then flows into the reaction compartment, which lead to non-equilibrium steady state conditions. Hence we now present a Michaelis-Menten reaction mechanism with continuous substrate inflow into a compartment (see, **FIG 2B**). The scheme describing this reaction mechanism is:

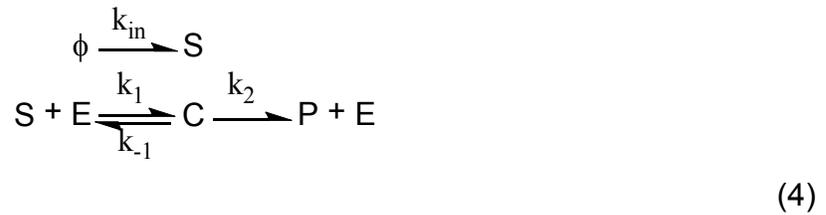

(4)

where $k_{in}$ is the substrate inflow (or production) rate. This reaction achieves steady-state when $k_{in}$ is less than the limiting rate of the reaction, $k_2 e_0$. At steady-state, the DREs can be solved exactly leading to an analogous expression to the Michaelis-Menten equation [58]:

$$\frac{dp}{dt} = k_{in} = \frac{k_2 e_0 s}{K_M + s}.$$

(5)

In this case the Michaelis-Menten equation provides a relationship between the steady-state rate of product formation, which equals $k_{in}$, and the steady-state substrate concentration. This result is mathematically the same as the Michaelis-Menten equation (2). The only difference is as follows. For the Michaelis-Menten reaction mechanism (1), the measurement of the Michaelis-Menten constant and the turnover number are estimated from initial rate experiments in quasi steady-state conditions. However, for the Michaelis-Menten reaction mechanism with substrate inflow (4), the kinetic constants are estimated from rate experiments in non-equilibrium steady-state conditions. We note that Eq. (5) assumes fluctuations are negligible, which is not typically the case in single molecule experiments. Generally fluctuations in chemical systems which are in a quasi-steady-state differ from those in a non-equilibrium steady-state [59]; this is the case for enzyme catalysed reactions as well, as we shall see next.

**Analysis of the Michaelis-Menten reaction mechanism with substrate inflow catalysed by few enzyme molecules**

Now we relax the condition of small fluctuations. Stefanini et al. [60] analysed the Michaelis-Menten reaction mechanism with substrate inflow (4) catalysed by one enzyme molecule (**FIG. 2B**) in a compartment, and found an exact analytical solution for the CME approach. They discovered that the relationship between the mean steady-state rate of product formation is not given by the Michaelis-Menten equation Eq. (2) but by a more complex expression. By explicitly showing the dependence of the propensities in the CME on the compartmental volume $\Omega$, the mean rate of product formation is (see Eq. (70) in [60]):

$$\frac{d\phi_P}{dt} = k_{in} = k_2 \frac{K_M + \phi_S}{2} \left( \sqrt{1 + \frac{4\phi_S'}{\Omega K_M (1+\phi_S')^2}} - 1 \right)$$

(6)

Note that $\phi_S$ is the substrate concentration for the CME which is typically different than $s$, the substrate concentration for the DREs. The same notation is used for the product concentration. The notation $\phi_S'$ refers to the non-dimensionalised concentration $\phi_S' = \phi_S / K_M$. Eq. (6) is to be contrasted with Eq. (5) which considered the same reaction mechanism (4), but neglecting fluctuations. Hence it is clear that if one tried to estimate the Michaelis-Menten constant and the turnover number from single molecule experimental measurements of the rate of product formation and steady-state substrate concentration using Eq. (5), then one could obtain misleading results for these constants. In contrast, use of Eq. (6) would lead to accurate results.

By an inspection of Eq. (6) it follows that in the limit $K_M \Omega \gg 1$, Eq. (6) reduces to the Michaelis-Menten equation (2) with $p = \phi_P, s = \phi_S$ and $e_0 = 1/\Omega$. Given the definition of $K_M$ and the fact that $k_{-1} + k_2$ represents the frequency with which complex dissociates and $k_1/\Omega$ the frequency with which a substrate and an enzyme molecule associate, it follows that $K_M \Omega \gg 1$ implies the condition whereby bimolecular binding occurs relatively rarely compared to complex breakdown. Hence fluctuations in the substrate concentration are small and the bimolecular nature of the reaction is diminished, i.e. the two key ingredients which are necessary to obtain discrepancies between the CME and DRE predictions for the mean concentrations [51], are missing. This reasoning explains why the stochastic model leads to the deterministic Michaelis-Menten equation in the limit of large $K_M \Omega$. This result is as well consistent with the derivation of a single molecule Michaelis-Menten equation by Kou et al. [52] and Qian [53] under the assumption of a constant non-fluctuating number of substrate molecules. In the current example deviations from the Michaelis-Menten equation are due to substrate fluctuations; deviations are similarly possible due to fluctuating $K_M$ which model enzyme conformational dynamics [61]. Deviations from the DRE predictions of the reversible Michaelis-Menten reaction mechanism with one enzyme molecule and in a non-equilibrium steady state, have also been investigated by Darvey and Staff [62] and Qian and Elson [63].

Conversely deviations from the Michaelis-Menten equation due to substrate fluctuations become significant for single enzymes confined in small volumes. For example, for a single enzyme with a $K_M$ between 1 and $10^4$ μM (a range reported for physiological conditions [64]), the critical volumes below which deviations are important are $\Omega = K_M^{-1} \approx 10^{-25} - 10^{-21} m^3$ which roughly corresponds to a cubic compartment with a side length in the range 5 - 100 nanometres. Hence the fluctuation-induced deviations from the Michaelis-Menten equation, as described by Eq. (6), are important for single enzyme molecules in small compartments with diameter of roughly 100 nanometres, such as carboxysomes [65], and bacterial microcompartments [66]. On the other hand the deviations would be insignificant for a single enzyme in a nucleus since the latter is typically micron sized or larger.

Thus far we discussed the case of a single enzyme molecule in a compartment. As previously remarked this is useful as a means to estimate the maximum deviations expected from the predictions of the deterministic approach. The predictions from this single molecule approach are reflective of the multi-enzyme case whenever conditions are such that different enzyme molecules carry out catalysis independent of each other. Such conditions naturally follow when the substrate is consumed slowly, in which case both the single and multi-enzyme dynamics follow the Michaelis-Menten equation. However it has not been possible to obtain an exact analytical expression for the probability distribution solution of the CME of the Michaelis-Menten reaction mechanism with substrate inflow (4) catalysed by many enzyme molecules for general rate constant values. Stochastic simulations of reaction mechanism (4) with enzyme molecule numbers in the range of 10-100 and with physiologically realistic parameters show that whenever the criterion $K_M \Omega \ll 1$ is satisfied, the Michaelis-Menten equation Eq. (5) does not accurately describe the relationship between the rate of product formation and the mean substrate concentration [58].

It has been recently shown [58] that to a good degree of approximation, the aforementioned relationship is quantitatively well described by the following equation:

$$\alpha + \left(1 - \frac{\phi_S}{K_M + \phi_S}\right) \frac{\alpha^2}{\Omega(K_M + e_0(1-\alpha)^2)} = 1 - \frac{\phi_S}{K_M + \phi_S},$$
(7)

where $\alpha$ is the mean rate of product formation normalized by the limiting rate: $k_{in}/k_2 e_0$. Equation (7) has been derived using a novel type of rate equation called Effective Mesoscopic Rate Equations (EMREs) [40], which approximate the mean concentrations predicted by the CME and which reduce to the DREs in the limit of large molecule numbers. Whereas DREs are derived from the CME by assuming zero fluctuations, the EMREs are derived by assuming small but non-zero fluctuations. This implies that the DRE predictions do not take into account the coupling between the mean concentrations and the covariance of fluctuations inherent in the CME approach, whereas the EMRE does preserve such coupling, albeit in an approximate sense. This implies that the EMRE approach presents a more accurate means of predicting mean concentrations; indeed EMREs have been shown to closely match the CME for molecule numbers greater than a few tens (see next section). Equation (7) thus provides an accurate means to estimate the Michaelis-Menten constant and turnover number from single-cell measurements of the mean substrate concentration and the mean rate of product formation for reaction mechanism (4).

## Stochastic analysis of a Michaelis-Menten reaction mechanism coupled to complex substrate inflow

In the last section we have considered the Michaelis-Menten reaction mechanism with substrate inflow. This model captures the basic phenomenon of substrate input but lacks biochemical detail. Now we consider a more complex reaction mechanism of substrate inflow, which has been recently used to model the transcription, translation and degradation of a substrate in *E. coli* [67]:

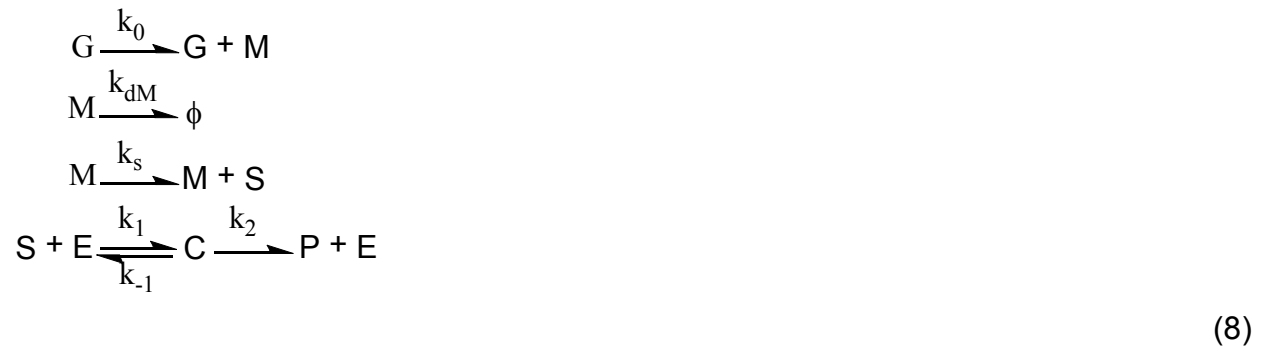

(8)

In the above reaction mechanism, G can be considered a gene coding for the substrate and M is its mRNA. $k_0$ is the transcription rate and $k_S$ is the translation rate. It is assumed that the gene G has only one copy in the cell. The translated protein S is then catalysed by an enzyme E to a final product P via a single complex intermediate C. A simple ubiquitous example of this reaction mechanism is the degradation of a translated protein S into a non-active form P. The kinetics of such a process has been shown to follow Michaelis-Menten kinetics [68] and hence the use of the Michaelis-Menten reaction mechanism as a very simple model of the intricate underlying degradation machinery. Our reaction scheme (8) can be seen as a refinement of the standard model of gene expression in *E. coli* [69, 70] in which substrate degradation is modelled via a first-order reaction.

The DRE model for the reaction mechanism (8) in non-equilibrium steady-state conditions can be described with an analogous expression to the Michaelis-Menten equation:

$$\frac{dp}{dt} = \frac{k_s k_0 g}{k_{dM}} = \frac{k_2 e_0 s}{K_M + s}.$$

(9)

Note that the quantity $g$ is the gene concentration. Thus, deterministically, i.e., in the absence of fluctuations, we again have a Michaelis-Menten relationship between the rate of product formation and the mean substrate concentration, as previously found for the simpler model in the previous section. The single and many enzyme copy number versions of this model cannot be solved analytically. In **FIG. 4** we compare the numerical predictions of the two approaches for parameters $k_0 g = 0.024 \min^{-1}$, $k_{dM} = 0.2 \min^{-1}$, $k_S = 1.5 \min^{-1}$, $k_{-1} = k_2 = 2 \min^{-1}$, and $k_1 = 400 \, (\mu M \min)^{-1}$. The enzyme copy numbers were fixed to 60 in a volume equal to the average volume of an *E. coli* cell. Note that the relative percentage difference between the CME's and the DRE's prediction of the mean substrate concentration in steady-state conditions is close to 100%. This is considerable, which highlights the breakdown of the DRE approach to modelling enzyme catalysed reactions with low molecule numbers.

The difference between reaction mechanisms (4) and (8) stems from the breakdown of the input reaction from one reaction step in (4) to two reaction steps in (8). Hence the inclusion of the intermediate mRNA production step could be the culprit for the unexpectedly large deviations from the Michaelis-Menten equation. Now, it is known that under certain conditions the mRNA step leads to substrate molecules being produced in large bursts at random times. These conditions occur when the lifetime of mRNA is much shorter than that of proteins which is typical in bacteria and yeast [71] (and *in vivo* measurements of protein expression verify that protein expression can occur in sharp bursts [72, 73]). What this means is that for short periods of time after a burst occurs, there can be much more substrate than the enzyme can catalyse, even if working at maximum speed. Consequently, substrate accumulates. The CME captures these random bursts whereas the DRE does not, which explains why the DRE underestimates the substrate concentrations in **FIG. 4**. Generally it has been shown that for the Michaelis-Menten reaction mechanism with substrate inflow occurring in bursts at a given $K_M\Omega$, the deviations from the deterministic Michaelis-Menten equation will be larger than those for the Michaelis-Menten reaction mechanism with substrate inflow (no bursts) at the same $K_M\Omega$ [58].

As we illustrate in **FIG. 4**, the CME and the DREs predict numerically different time courses for the same set of parameters. This implies that the estimation of rate constants from time course data of single cells would also lead to different numerical estimates between the CME and the DREs. In **FIG. 4**, we also illustrate the closeness of the EMRE prediction to that of the CME for the reaction mechanism (8). It is a considerable improvement over the DRE approach. Hence we expect that parameter estimation could be carried out effectively using EMREs instead of DREs for enzyme-catalysed and other biopolymer-mediated reactions in stochastic conditions.

## Conclusions

We have briefly summarised the state of the field of stochastic enzyme kinetics for the single substrate, single enzyme Michaelis-Menten reaction mechanism. While the foundations of the field were laid over 50 years ago, many significant theoretical challenges have only been surmounted in the past decade. These developments are spurred in large part by technological advances enabling us to probe the kinetics of single molecule reactions on nanometre length scales which are relevant to understanding kinetics at the cellular level and inside artificial nanoscale compartments [74] and biomimetic reactors [75]. We note that while recent experiments have validated some of the theoretical results for single molecules with no substrate inflow, thus far experimental validation of theoretical results for enzyme systems with substrate inflow has been lacking; hence this field still presents many challenges to be solved.

In this review, we present two take home messages:

(i) The CME (stochastic) and DREs (deterministic) approaches may predict different numerical values for the mean substrate, enzyme and complex concentrations in time, as well as different steady-state concentrations for a given set of rate constants. These differences are typically small for the Michaelis-Menten reaction mechanism but significant for the Michaelis-Menten reaction mechanism with substrate inflow. The differences increase with decreasing $K_M \Omega$ and are particularly conspicuous when substrate inflow occurs in bursts.

(ii) Besides providing accurate predictions of the mean concentrations, the CME approach also provides additional information regarding the fluctuations about these concentrations and in particular the probability distribution of the waiting time between successive product turnover events. The latter could be used to distinguish between rival models of enzyme action.

Point (i) has important implications for the estimation of rate constants of enzyme-catalysed and other biopolymer-mediated reactions. Estimated rate constants can differ significantly depending on the approach (CME or DREs) adopted to model the reaction. The CME is superior since it is valid for both reactions occurring with large or small molecule numbers. Unfortunately, the estimation of rate constants from stochastic simulations of the CME is highly time consuming and has only started to be tackled quite recently [76]. The EMRE approach may present a way around this challenge since parameter estimation methods are well developed for rate equations [77]. These approaches have thus far been exclusively used with DREs but can also be used with EMREs since the latter are also a type of rate equation.

Point (ii) has important implications for the development of novel experimental approaches, which can probe fluctuations in single molecule events at fine temporal resolution [29]. The CME can then be used with this data to infer a wealth of information about the reaction dynamics, which cannot be accessed through DREs.

The future of stochastic enzyme kinetics lies in the development of experimental techniques to access real-time enzyme-catalysed and other biopolymer-mediated reactions at the single molecule level inside living cells. In parallel, it is also essential to develop novel theoretical toolkits so that we can infer reaction mechanisms and estimate rate constants from the emerging single-cell high-resolution data.

## Acknowledgements

We would like to thank Philipp Thomas (University of Edinburgh), Roberto Miguez and Caroline Adams (University of Michigan) for carefully reading the manuscript. This work has been partially supported by the James S. McDonnell Foundation under the 21st Century Science Initiative, Studying Complex Systems Program.

**Figure legends**

**FIG. 1. A single molecule fluorescence microscope can read out the turnover of single immobilized enzyme molecules as they convert fluorogenic substrate in solution into fluorescent product, often in bursts of activity.** (A) Schematic illustration of objective-type TIRF microscope. (B) Real-time single-molecule recordings of enzymatic turnovers as fluorogenic substrate is converted into fluorescent product. Each emission intensity peak corresponds to an enzymatic turnover.

**FIG. 2. Schematic illustration of the two cases primarily treated in this review.** (A) The Michaelis-Menten reaction with one enzyme molecule and in a closed compartment. (B) The Michaelis-Menten reaction with one enzyme molecule and with substrate inflow. The latter could for example model unidirectional active transport of substrate to a compartment or else the production of substrate by an upstream process.

**FIG. 3. Differences between the DRE and CME predictions of the mean concentrations of enzyme and substrate for the Michaelis-Menten reaction catalysed by a single enzyme molecule.** (A) Reproduces a case first studied in [45] where initially there is a single molecule of substrate and the parameters are $k_1/\Omega = 10, k_{-1} = 2, k_2 = 1$. (B) Parameters are kept as in the previous case but the initial number of substrate molecules is increased to 5. Note that the discrepancies observed between the CME and the DRE approaches are only significant for very low numbers of substrate molecules. Time is in non-dimensional units.

**FIG. 4. Theoretical discrepancy between the stochastic and deterministic approaches in a gene expression model involving enzyme catalysis.** The model considers gene expression of substrate and its subsequent catalysis into product via the Michaelis-Menten reaction mechanism according to the scheme (8). The cell volume is a femtolitre, which is on the range of the volume of an *E. coli*. The total number of enzyme molecules is 60 (see text for the rest of the parameters). The initial conditions are such that there is no substrate, mRNA and product and that the free enzyme

concentration equals the total enzyme concentration. (A) The deterministic rate equation (DRE, dashed line) severely underestimates the mean concentration prediction of the stochastic simulation algorithm (SSA, red line) while the Effective Mesoscopic Rate Equation (EMRE, black line) provides a much better approximation to the latter. (B) Whereas the DRE approach assumes a probability distribution of substrate molecules, which is very sharp, i.e., no fluctuations, in contrast the actual probability distribution of substrate molecules (in steady-state conditions), as obtained using the SSA has a very slowly decaying tail. The mean concentration predicted by the DRE is closer to the mode of the distribution than to its average (see [78] for a detailed discussion of this phenomenon).

**Figures**

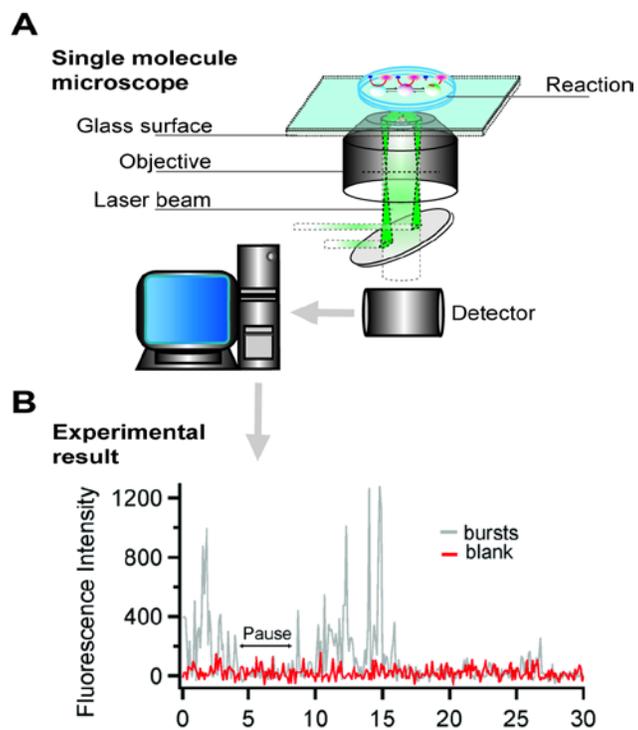

FIG. 1. Grima et al. (2013)

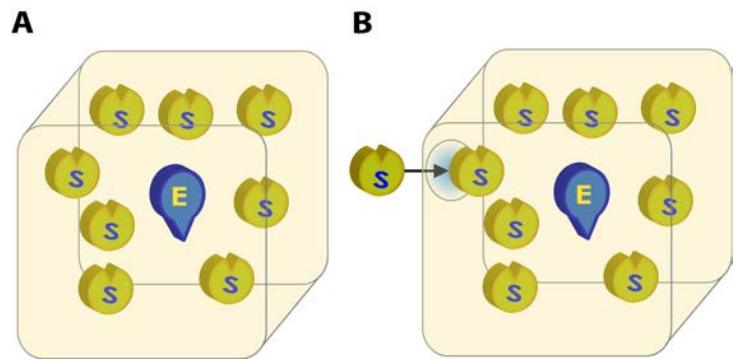

**Fig. 2.** Grima et al. (2013)

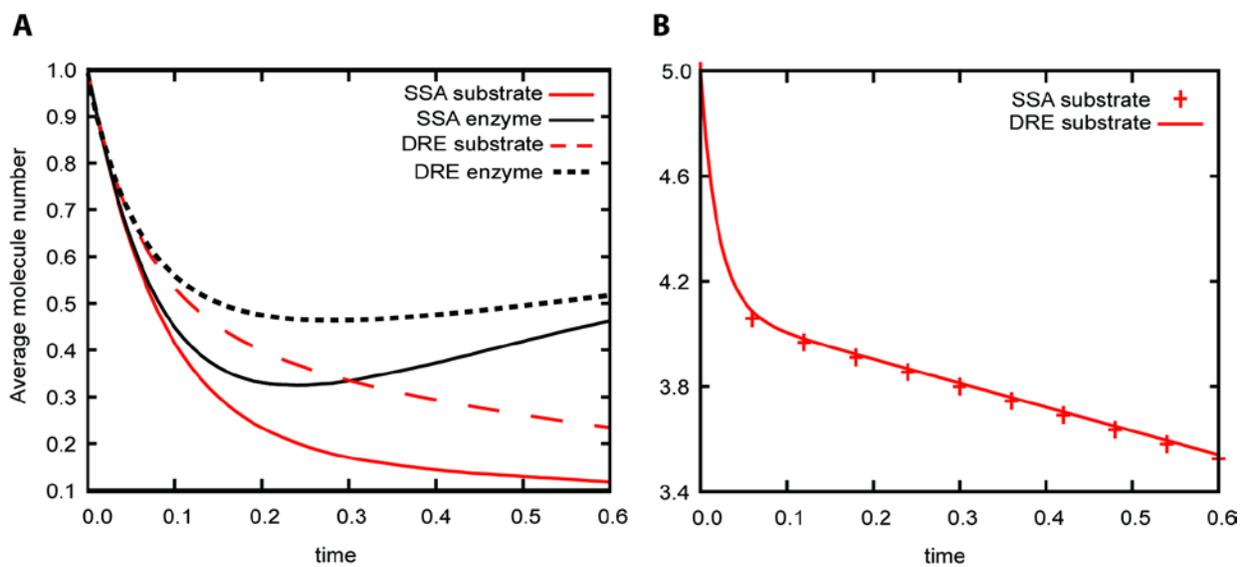

**Fig. 3.** Grima et al. (2013)

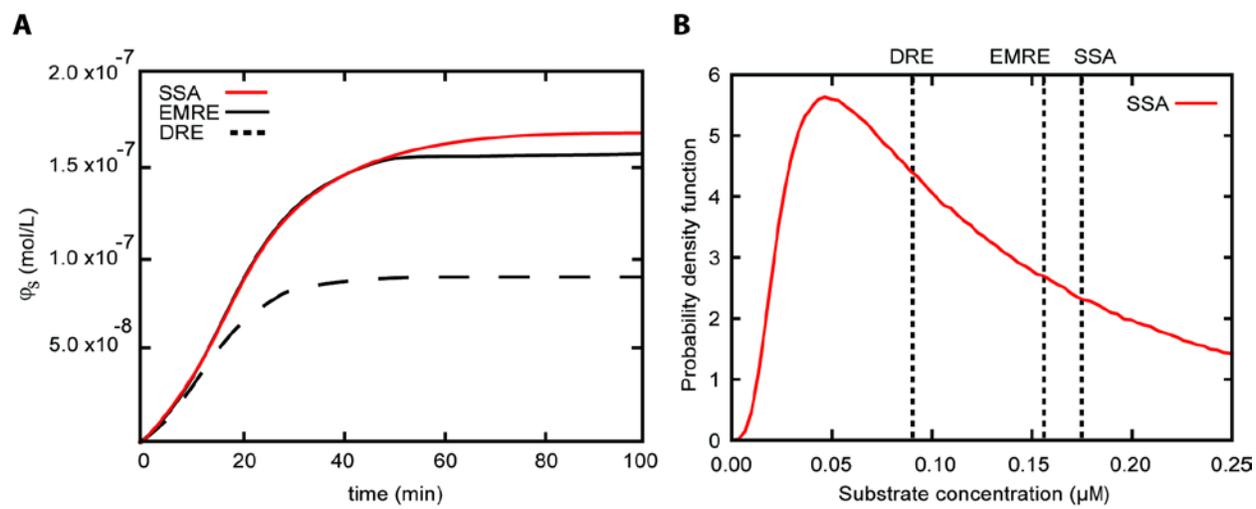

**Fig. 4.** Grima et al. (2013)